\documentclass[prl,twocolumn,superscriptaddress]{revtex4-1}

\pdfoutput=1
\usepackage{epsfig,graphicx,amsbsy,amssymb,latexsym,amsfonts,amsmath}
\usepackage{pstricks}

\newcommand {\be}{\begin{equation}}
\newcommand {\ee}{\end{equation}}

\newcommand{\C}{{\mathbb C}}
\newcommand{\e}{\epsilon}

\newcommand{\het}{\text{het}}
 \newcommand{\nn}{\nonumber}

\newcommand{\tr}{{\rm tr\,}}

\newcommand{\vt}{\vartheta_1}
\newcommand{\Z}{{\mathbb Z}}

\begin{document}

\title{Elliptic Genus, Anomaly Cancellation
and Heterotic M-theory}
\author{Kang-Sin Choi} 
\email{kangsin@ewha.ac.kr}
\affiliation{Scranton Honors Program, Ewha Womans University, Seoul 03760 \rm KOREA} 
\author{Soo-Jong Rey}
\email{sjrey@snu.ac.kr}
\affiliation{School of Physics and Astronomy, Seoul National University, Seoul 08826 \rm KOREA}

\begin{abstract}
We derive global consistency condition for strongly coupled heterotic string in the presence of M5-branes. Its elliptic genus is interpretable as generating functional of anomaly polynomials and so, on anomaly-free vacua, the genus is both holomorphic and modular invariant. In holomorphic basis, we identify the modular properties  by calculating the phase. By interpreting the refinement parameters as background curvature of tangent and vector bundles, we identify the extended Bianchi identity for Kalb--Ramond field of heterotic M-theory in the presence of arbitrary numbers of M5-branes.
\end{abstract}

\maketitle 
The anomaly of gauge and flavor symmetries played an important role as a unique window to physics at short-distances, encompassing neutral pion decay, baryon number violation, matter contents of the Standard Model, quantum Hall edge states, topological insulators, and so on.   
For chiral gauge theories, anomaly {\em structure} is elegantly organized,  while anomalous {\em field contents} lead to quantum inconsistency. It restricts possible consistent vacua of the theory purely in terms of low-energy degrees of freedom, i.e. spectrum of massless fields. 
At high-energy scale, the anomaly structure is embedded to {\em global consistency condition} of ultraviolet completion such as string theory: modular invariance of closed strings and tadpole cancellation of open strings. These conditions, which lead to a specific form of anomalies, can be read off from the partition functions and their behavior in the complex plane. One expects anomaly structure severely constrains the functional form of partition function.

In this Letter, we study global consistency condition of heterotic M-theory \cite{Horava:1995qa, Kim:1997uv} or strongly coupled heterotic string theory \cite{Rey:1997hj}, whose new feature is the presence of M5-branes in the M-theory bulk. The elliptic genus \cite{Oc,Witten:1986bf}  
is known to be the generating function for anomaly polynomial, and tells us that anomaly cancellation occurs when it is both holomorphic and modular invariant \cite{Schellekens:1986yi,Schellekens:1986yj}. The fluctuations of M5-branes are described by M2-branes stretched between M5-branes. Thus, to identify consistent vacua in the presence of M5-branes, we may analyze the corresponding elliptic genera as a proxy for anomaly cancellation. Such elliptic genera are most elegantly computed by the topological vertex formalism \cite{Aganagic:2003db,Iqbal:2007ii}.  

\paragraph{Elliptic genus}
The elliptic genus is defined by trace over the Ramond sector of heterotic string worldsheet with ${\cal N} = (0,2)$ supersymmetry \cite{Witten:1986bf} 
\be \label{ellgenus}
 Z(q, {\bf x})  =  {\rm Tr\,}_{\rm R} q^{H} {\bar q}^{\overline H} (-1)^{F} \prod_a x_a^{{Q}_a}.  
\ee
Here, $q= e^{2 \pi i \tau}$ with $\tau$ the modular parameter of torus, $F$ is the fermion number, $Q_a$ are the set of global charges, and $H, (\overline H)$ are the (anti)holomorphic Hamiltonians. For compact target space, their spectra are discrete, rendering the sum over states well defined. The ${\cal N} = (0,2)$ supersymmetry ensures that the elliptic genus is independent of $\bar q$. 

For non-compact target space, the elliptic genus is afflicted by infrared divergence due to infinite target space volume. 
To avoid it, we regularize it so that the heterotic string is localized at a point. We do so by introducing $\Omega$-deformation and extract anomaly structure from the associated global symmetry. It is formally analogous to orbifolding the ambient target space. Denoting transverse 8 coordinates as $z_m \equiv x_{2m} + i x_{2m+1} , m=1,2,3,4,$ we set the $\Omega$-deformation of the target space $\C^4$ \cite{Nekrasov:2002qd} by twisting
\be \label{twisting}
 z_m \to e^{2\pi i \e_m} z_m, \quad m=1,2,3,4,
\ee
and by simultaneously shifting the vectors $\vec p = \vec p_L \oplus \vec p_R$ in the internal $E_8\times E_8$ lattice
 $\Gamma_{8}\oplus \Gamma_8$
\be \label{shifting}
 \vec p_A \to \vec p_A + \vec m_A, \qquad (A=L,R)
 \ee
where $\vec m_A$ is {an} 8-component vector in $\Gamma_{8}$. 

We further compactify longitudinal $x^0, x^1$ directions on a torus. The $\Omega$-deformation has the effect that, whenever we go around the cycles of the torus, we have the above twisting (\ref{twisting}) and (\ref{shifting}).
One-eighth of 32 supercharges survive if $\e_1+\e_2+\e_3+\e_4=0$. Supercharges are further reduced in the presence of M9 and M5 branes.

In the case of weakly coupled heterotic string, viz. no M5 branes in the bulk, we obtain the corresponding elliptic genera using the Hirzebruch--Riemann--Roch index theorem \cite{Schellekens:1986yj,Witten:1986bf}. For a single heterotic string, we can decompose the elliptic genus as 
\be \label{anomalypolyn}
   Z_1^{\rm het}= \frac{1}{16 \pi^4 }  \hat A(R) P(\tau,F) P_B(\tau,R) \, {\rm vol}\,(\C^4),
\ee
where, under the $\Omega$-deformation (\ref{twisting}), 
\begin{align}
 \hat A(R) &\equiv \prod_{j=1}^4 \frac{\pi \e_j}{\sin (\pi \e_j)}  , \label{Diracgenus} \\
 P(q,F)&\equiv \frac{A_1 (\vec m_L) A_1 (\vec m_R)}{\eta(\tau)^{16}} , \label{gaugeshift}\\
  P_B(q,R) &\equiv \prod_{j=1}^4  \frac{2\sin (\pi \e_j) \eta(\tau)}{\vt(\e_j)}  , \label{spacetwist} \\
 {\rm vol}\, \C^4 &=  \frac{1}{\e_1 \e_2 \e_3 \e_4}  \label{volC4}. 
\end{align}
The Dirac genus (\ref{Diracgenus}) counts the `the number of fixed points' under the twisting (\ref{twisting}). The next factor $P(q,F)$ 
is the generating function for Chern character $A_1 (\vec m_L) A_1 (\vec m_R) \equiv \sum_{\vec p \in  \Gamma_{8}\oplus \Gamma_8} q^{\vec p^2/2} e^{i p \cdot \vec m}$ that comes from the lattice  $\Gamma_{8}\oplus \Gamma_8$. A similar spacetime twist is encoded in $P_B(q,R)$ in (\ref{spacetwist}). As $\e_i$ is the eigenvalue of the Riemann curvature tensor in the $i$-th direction in (\ref{twisting}) \cite{Nekrasov:2002qd}, Eq.(\ref{volC4}) corresponds to the regularized volume localized at the fixed point.

As recalled, the elliptic genus is the generating function for anomaly polynomials  \cite{Schellekens:1986yj}. In general, however, an elliptic genus (\ref{ellgenus}) is not invariant under the modular transformation. It is invariant under the shift $T:\tau \to \tau+1$ of modular transformation $SL(2,\Z)$, but not under the inversion $S:\tau \to -1/\tau$ \cite{Schellekens:1986yi} .  It was found that \cite{Schellekens:1986yj}, regarding $\e_i$ and $m_I$ as eigenvalues of Riemann curvatures and field strengths of Cartan subalgebra of $E_8\times E_8$, the phase under $S$ transformation, 
\be \label{modularanomaly}
Z^{\rm het}_1 (-1/{\tau})  =Z^{\rm het}_1(\tau) \exp\left[\frac{\pi i}{\tau} ( \tr R \wedge R - \tr F \wedge F) \right], 
\ee
 reveals the Bianchi identity for $H$, the field strength of Kalb--Ramond field $B$. 
We can interpret the parameters in the elliptic genus (\ref{anomalypolyn}) as skew-eigenvalues of the vector bundles and the tangent bundles
\be 
 \tr F \wedge F =  \sum_{I=1}^{16} m_I^2, \quad 
 \tr R \wedge R =  \sum_{m=1}^4 \epsilon_m^2, 
\ee
where $F$ is the field strength of the $E_8 \times E_8$ and $R$ is Riemann curvature tensor. The latter agrees with the relationship between the curvature and the volume (\ref{volC4}).

\paragraph{M5- and M9-branes}

We describe strongly coupled heterotic string theory by M-theory compactified on an interval \cite{Horava:1996ma}. We have two M9-branes with $E_8$ gauge theories at the ends of the interval in, say, the $x^{10}$-direction, $0 \le x^{10} \le L_M$. 
An M2-brane stretched between two M9-branes gives rise to heterotic string
\cite{Kim:1997uv, Kabat:1997za}.

We may put additionally a number of M5-branes at various places in the interval, away from M9-branes. Their locations are $z^3=z^4=0$ with $x^{10}$ arbitrary within $[0, L_M]$. The setup gives rise to so-called M- and E-strings of variable tension, obtained from M2-branes connected between different M9/M5 branes \cite{Haghighat:2013gba,Haghighat:2014pva}.
M-strings come from M2-branes stretched between two M5-branes, and describe inter-brane fluctuations \cite{Haghighat:2013gba,Haghighat:2014pva}. 
E-strings come from M2-branes stretched between M9- and M5-branes, and describe fluctuation of M5-brane relative to the M9-brane.  

The elliptic genus corresponding to this setup again contains information on anomaly structure and hence on global consistency conditions, but now including new contributions from M5-branes. Their presence is a source of technical as well as conceptual complications but, as we show momentarily, the new elliptic genus can still be computed for arbitrary number of M5-branes and heterotic strings.  To probe non-Abelian structure of M- and E-strings, one would need to uplift to the F-theory dual, as analyzed in \cite{Choi:2017vtd}.

More specifically, the presence of M5-branes affect the modular transform (\ref{modularanomaly}), modifying the anomaly structure. In this Letter, we extract this information from the corresponding elliptic genus, which we calculate from the refined topological vertex method \cite{Aganagic:2003db,Iqbal:2007ii}.

\paragraph{Elliptic genus from refined topological vertex}

We can calculate elliptic genera of non-perturbative heterotic string in the presence of $n$ M5-branes, using the refined topological vertex method. The calculation boils down to the product of defect operators
\be \label{HetElpGen} \begin{split}
   Z_n (\tau,\vec \e,\vec m)= & \sum_{\{\nu_a\}}D^{M9}_{L,\nu_1} \left( \prod_{a=1}^n D^{M5}_{\nu^t_a \nu_{a+1}} \right) D^{M9}_{\nu_{n+1}^t,R} \\
+&\sum_{\{\nu_a\}}D^{M9,c}_{L,\nu_1} \left( \prod_{a=1}^n D^{M5}_{\nu^t_a \nu_{a+1}} \right) D^{M9,c}_{\nu_{n+1}^t,R} .
\end{split}
\ee
The defect operator $D^{M5}_{\nu_a^t \nu_{a+1}}$ for $a$-th M5-brane connected by M2-branes with the tableaux $\nu_a$ on the left and $\nu_{a+1}$ on the right was computed in Ref. \cite{Haghighat:2013gba}. Here, we take the convention that a Young tableau $ \lambda$ encodes the configuration of M2-branes by descending ordered set of numbers $\lambda=(\lambda_1,\lambda_2,\dots), \lambda_1 \ge \lambda_2 \ge \dots.$ The superscript $t$ refers to transpose. The size of $\lambda$ is $|\lambda|=\sum_i \lambda_i = \sum_j \lambda^t_j$.
 
We also have two operators for M9-branes, $D^{M9}_{L,\nu_1}$ and $D^{M9}_{\nu_{n+1}^t,R}$. For our foregoing analysis, however, we do not need detailed form of them (they can be found in \cite{Haghighat:2014pva,Kim:2014dza,Cai:2014vka}) except for the followings. { First, we have exchange symmetry $ D^{M9}_{L,\nu^t} = D^{M9}_{L,\nu} (\e_1 \leftrightarrow \e_2)$. Second, operationally, these defect operators are obtainable from the elliptic genus of E-strings by assuming that M5-branes are located at $(z^3,z^4)=(0,0)$. This choice, however, explicitly breaks the $SO(8)$ symmetry of M9-brane worldvolume. To restore $SO(8)$, we may symmetrize the orientation of M5-brane worldvolume. Equivalently, we may fix the M5-branes orientation as above and then symmetrize M9-brane worldvolume coordinates $(z_1, z_2, z_3, z_4)$. The net effect is to introduce additional defect operators $D^{M9,c}_{L,\nu} \equiv D^{M9}_{L,\nu} (\e_1 \leftrightarrow \e_3)$, which implies $D^{M9,c}_{L,\nu^t} \equiv D^{M9}_{L,\nu} (\e_1 \leftrightarrow \e_4)$ \cite{Haghighat:2014pva}.}
This is how we expressed the partition function in the form (\ref{HetElpGen}).

\paragraph{Modular anomaly and holomorphic anomaly}
A consistent field content must give rise to modular invariant and holomorphic elliptic genus. In general, it is {\em not} possible to maintain {\em both} of them. Basic building block of elliptic genera is the Jacobi $\vartheta$-function $\vartheta_1$.
We can check that $\vartheta_1 (\frac{a \tau +b }{c\tau+d}; \frac{z}{c \tau+d}) = (c\tau+d)^{1/2} e^{\pi i z^2/(c\tau+d)} \vartheta_1(\tau;z).$
We can understand the reason why the phase is quadratic in $z$.
Expanding it, 
$$
 \vartheta_1(z) = \eta(\tau)^3 (2\pi i z) \exp \left ( \sum_{k=1}^\infty \frac{B_{2k}}{(2k)(2k)!} E_{2k} (2 \pi i z)^{2k} \right),
$$
where $E_{2k}$ are $2k^{\rm th}$ Eisenstein series and $B_{2k}$ are Bernoulli numbers. 
All the $E_{2k}$ for $k\ge 2$ are holomorphic modular form and generated by $E_4$ and $E_6$. The exception is  $E_2$ which transforms under $SL(2,\Z)$ as
$$ E_2 \left( \frac{a\tau+b}{c \tau+d} \right) = (c\tau+d)^2 E_2 (\tau) - \frac{6ci}{\pi} (c\tau+d), $$
where $a,b,c,d \in\Z, ad-bc=1$.
We may redefine this to be modular at the price of giving up holomorphy,
$$ \hat E_2 (\tau, \bar \tau) = E_2 (\tau) - \frac{6i}{\pi (\tau-\bar \tau)}, $$
such that 
$ \hat E_2 \left( \frac{a\tau+b}{c \tau+d} , \frac{a \bar \tau+b}{c \bar \tau+d}\right) = (c\tau+d)^2 \hat E_2 (\tau,\bar \tau).$
Thus, anomalous phase of the elliptic genus (\ref{HetElpGen}) is only up to quadratic because the only non-holomorphic part in $\vartheta_1$ is the coefficient of $E_2$:
\be
\frac{30}{\pi^2} \frac{\delta \log   Z_n}{\delta E_2} 
.
\ee

If the phase vanishes for a given field content, anomaly cancellation is ensured for the corresponding vacuum. 
For generic $\vec \e$ and $\vec m$, we have non-invariant phase under $S$ for the $Z_n(E_2, E_4, \cdots)$ in the holomorphic basis.
Although the complete expression for (\ref{HetElpGen}) is unknown, for extracting information on anomalies, it suffices to study the phases under the $S$ modular transformation. Being additive, we separate the phase of each term of (\ref{HetElpGen}) into two separate pieces. 

First, the transformation of M9 defect operators in (\ref{HetElpGen}) is {the same as that of} the elliptic genus of weakly coupled $k$ heterotic strings. The latter can be obtained by Hecke transformation of single string \cite{Dijkgraaf:1996xw, Haghighat:2014pva}. We found that this is a modular form provided $\nu_1=\nu_{n+1}$, so that 
\be \label{heteroticHecke}
\begin{split}
  Z^{\het}_k \equiv
 & \sum_{|\nu_1|=k}D^{M9}_{L,\nu_1}  D^{M9}_{\nu_1^t,R} 
+\sum_{|\nu_1|=k}D^{M9,c}_{L,\nu_1} D^{M9,c}_{\nu_1^t,R}  \\
= &
 \frac{1}{k} \sum_{a,d>0} \sum_{b(\text{mod}\, d)}   Z_1^{\het} \left(\frac{a\tau+b}{d},a\vec \e,a\vec m \right), 
\end{split}
\ee
where the sum is over positive $a,d$ satisfying $ad=k$. Here $   Z^{\het}_1$ is the elliptic genus of a single heterotic string (\ref{anomalypolyn}).
It is not invariant under $S$ by a phase factor,
\be \label{hetmodular} 
  Z^{\rm het}_k (-1/{\tau}) 
=   Z^{\rm het}_k(\tau) \exp \left[\frac{ \pi i k}{\tau} \left(\sum_{i=1}^4 \epsilon_i^2 - \sum_{I=1}^{16} m^2_I \right)\right].
\ee

Second, the $S$ modular transformation of M5 defect operators in (\ref{HetElpGen}), $ \prod_{a=1}^n D^{M5}_{\nu^t_a \nu_{a+1}} \equiv {\cal D},$ is again a quasi-modular form provided $\nu_1=\nu_{n+1}$, while each individual factor is not. This can be seen from the relation between quantum dilogarithmic function and Jacobi $\vartheta$-function \cite{Haghighat:2013gba}. 

The phase of ${\cal D}^2$ is equal to the phase of $\prod_{a=1}^n D^{M5}_{\nu^t_a \nu_{a+1}} D^{M5}_{\nu^t_{a+1} \nu_{a}}$. Each factor
\be
D_{\nu^{t}\mu}^{M5} D_{\mu^{t}\nu}^{M5} =
\prod_{(i,j)\in\nu}
\frac{ \vartheta^{\e_2 +\e_3}_{ij,\nu\mu}  \vartheta^{\e_2+\e_4}_{ij,\nu\mu}  }
{ \vartheta^{\e_2}_{ij,\nu\nu}  \vartheta^{-\e_1}_{ij,\nu\nu}  }   \label{M5defect}  
\prod_{(k,l)\in\mu} \frac{ \vartheta^{\e_2+\e_3}_{kl,\mu\nu}  \vartheta^{\e_2+\e_4}_{kl,\mu\nu}  }
{ \vartheta^{\e_2}_{kl, \mu\mu}  \vartheta^{-\e_1}_{kl ,\mu\mu}  }  
\ee
where 
\be
 \vartheta^{\e}_{ij,\nu\mu} = \vartheta_1\big(\e-\e_1 (\nu_i-j)+\e_2(\mu^t_j-i)\big)
\ee
is quasi-modular form and only dependent on the type of tableaux $\nu$ and $\mu$  (We neglect overall phase which does not affect the phase change under $S$) . 

Utilizing Young tableaux identities, we found that the phase of ${\cal D}$ under the $S$ transformation is given by
\begin{align}
 &\prod_{a=1}^n  D^{M5}_{\nu^t_a \nu_{a+1}} \left(-1/{\tau}\right) =   \prod_{a=1}^n  D^{M5}_{\nu^t_a \nu_{a+1}} (\tau)  
 \label{thephase}   \\
 &\times  \exp \left[  \frac{ \pi i}{\tau}\left((|\nu_a|-|\nu_{a+1}|)^2\e_1\e_2 -(|\nu_a|+|\nu_{a+1}|)  \e_3 \e_4 \right) \right]. \nn
\end{align}
It is remarkable that, despite stack of M-strings {\em cannot} be understood as Hecke transform of a single M-string, the net phase depends only on the sizes of tableaux $|\nu_a|$, but not on their shapes.
For instance, in the case of two M5-branes ($n=2$) with $\nu_1=\emptyset=\nu_3, |\nu_2|\equiv k$, we have  the overall phase $ k^2 \e_1 \e_2 -  k \e_3 \e_4$ in unit of $\pi/\tau$. Previously, this was derived from the holomorphic anomaly equation of M-strings \cite{Haghighat:2013gba,Kim:2016foj,Shimizu:2016lbw,Choi:2017vtd} .

Hereafter, we require the coefficient of $\e_1 \e_2$ to vanish, viz. $|\nu_a|=|\nu_{a+1}| \equiv k$ for all $a$. Physically, this amounts to forbidding any leakage of M2-brane charge on M5-brane worldvolume.  The M2-brane charge simply flows from $\nu_{a+1}$ to $\nu_a^t$ as a local process in $(z_1,z_2)$ space. Indeed, $k$strongly coupled heterotic strings chopped by M5-branes give rise to $k$ M-strings in each interval.
Under the $S$ transform, each M5 defect operator generates an equal amount of phase, so
\be \label{Mmodular}
 \prod_{a=1}^n D^{M5}_{\nu_a^t \nu_{a+1}} \left(-1/{\tau}\right) =  \prod_{a=1}^n D^{M5}_{\nu_a^t \nu_{a+1}} (\tau)  \times e^{ - \frac{ 2 \pi i }{\tau}k\e_3 \e_4 }.
\ee
Putting togehter, we achieve the modular invariance by demanding that the phase (\ref{Mmodular}) from M5-branes cancels off the phase (\ref{hetmodular}) from strongly coupled heterotic strings. It is straightforward to generalize this cancellation mechanism to include the contribution proportional to $\epsilon_1 \epsilon_2$ in $(z_1, z_2)$-space.

\paragraph{Orbifolded transverse space}

Consider a special case of $\Omega$-background $\epsilon_3=\frac1N,\epsilon_4=-\frac1N$. The transverse space $\C^2(z_3,z_4)$ is an orbifolded (conic) space. In this case, the space becomes noncompact K3, $X_{\rm loc} \equiv \C^2/\Z_N$. The vector $\vec m$ is understood as associated shift vector \cite{Nibbelink:2007pn}.
 
To relate the twist and shift vectors to tangent and gauge bundles, respectively, we resolve $X_{\rm loc} $ at $Z =\{(z_3,z_4)=(0,0)\}$.
For $\Z_2$ orbifold, it takes the form
\begin{eqnarray}
 \C^*: (z_3,z_4,x) &\to& (\lambda^{-1} z_3, \lambda^{-1} z_4, \lambda^2 x),
 \nonumber \\
 \hat X_{\rm loc} &=& (X_{\rm loc} - Z) / \C^*.
\end{eqnarray}
This introduces an exceptional divisor $E = \{x=0\}$ with self-intersection $E\cdot E=-2$, which is an ALE space $A_1$, on top of the ordinary divisors $D_i =\{ z_i=0\},i=3,4.$ These divisors satisfy linear equivalence relations \cite{Lust:2006zh} 
\be \label{linequiv}
 D_3 \sim D_4 \sim -\frac12 E.
\ee

\paragraph{Bianchi identity including M5-branes}

The result above catches only the local contribution at the singular locus $(z_3,z_4) = (0,0)$.
Each M5-brane sources Kalb-Ramond magnetic flux $H$. From localization, we have a physical interpretation of the volume
\be
 \int dz_3 dz_4 \e_3 \e_4 = 1,\ \  \text{ i.e. }  \ \ \ dz_3 dz_4 = \frac{1}{\e_3 \e_4}.
\ee
So, we interpret the  phase in Eq.(\ref{Mmodular}) as Dirac $\delta$-function
\be \label{Delta}
 \e_3 \e_4 = \delta^2( z_3) \delta^2(z_4 ) \equiv \delta^4 (z_{3,4}).
\ee
The M5-branes are located at the locus $(z_3,z_4)=(0,0)$.

Adding (\ref{hetmodular}) to (\ref{Mmodular}) and using  (\ref{Delta}), we obtain local consistency condition in the background of $n$ M5-branes
\be \label{BianchiB2}
 - \sum^n_{a=1} \delta^4(z_{3,4}) + \frac12 \sum_{m=1}^{4} \epsilon_m^2 - \frac12 \sum_{I=1}^{16} m_I^2 =0 .
\ee
Note that this holds for any $k$, the number of strongly coupled heterotic strings. This is the result of localization in noncompact $\Omega$-background. Moreover, interpreting the localization as a physical orbifolded space, we can obtain the global consistency condition for compact spaces as well. For example, K3 corresponds to the blow-up of $\mathbb{T}^4/\Z_2$. It has sixteen fixed points, each of which { is locally} given by $X_{\rm loc} =\C^2/\Z_2$. Blowing them up, we find that \cite{Nibbelink:2007pn}, using the splitting principle, 
\be
 \frac{1}{2} \int_{\rm \widehat X_{\rm loc}} \tr  R \wedge R = D_1 \cdot D_2 + D_1 \cdot E + D_2 \cdot E =  \frac{3}{2} .  \ \ \ \ \
\ee
In the blow-down limit of the exceptional divisor, this contribution is concentrated at the fixed points. Flat bulk space away from the fixed points does not contribute. Finally, we glue the sixteen fixed points and build the K3 surface, for which 
$
 \frac{1}{2} \int_{\rm K3} \tr R \wedge R = 16 \cdot \frac{3}{2} = 24.
$

We also embed the gauge bundle associated with the twist (\ref{twisting}) \cite{Nibbelink:2007pn}
\be
i F =  \sum_I m_I H^I D_3 \Longrightarrow \frac12 \int_{\hat X_{\rm loc}} \tr F \wedge F = \frac14 {\vec m}^2,
\ee
where $H^I (I=1,\dots,16)$ are elements of Cartan subalgebra of $E_8 \times E_8$ and we used the relation (\ref{linequiv}). In the absence of M5-branes or magnetic sources, $H$ is exact $dH=0$, but the M5-branes should provide the $\delta$-function contributions. For $n$ many, we have
\be 
 \int_{\rm K3} dH =  n =  -\frac12 \int_{\rm K3} F \wedge F + \frac12 \int_{\rm K3} R \wedge R = -4 {\vec m}^2 + 24, \nonumber
\ee
which is precisely the anomaly cancellation condition \cite{Sagnotti:1992qw}, which is the Bianchi identity for Kalb--Ramond field $B_2$ now in the presence of the M5-branes in the bulk. We can readily generalize it to other orbifolds \cite{Lust:2006zh,Nibbelink:2007pn}. 

We have obtained anomaly cancellation condition for arbitrary number of tensor multiplets in six-dimensional non-perturtabative heterotic string. The key idea behind our derivation is the requirement that the elliptic genus must satisfy modular invariance and holomorphy simultaneously. It would be interesting to generalize the analysis to orbifolded M-strings \cite{Haghighat:2013tka, Hohenegger:2013ala, Hohenegger:2015cba, Ahmed:2017hfr}
and also to classify all possible globally consistent string configurations.

\paragraph{Acknowledgements} We thank Jin-Beom Bae, Stefan Hohenegger,  Amer Iqbal, and Taro Kimura for useful discussions. The work of KSC was supported by the National Research Foundation of Korea funded by the Ministry of Education under NRF-2015R1D1A1A01059940.

\end{document}